\documentclass[journal=jacsat,manuscript=article]{achemso}

\usepackage[version=3]{mhchem} 

\author{Kirtan P. Dixit}
\email{kirtankumar.p.dixit@nasa.gov}

\affiliation[NASA Marshall Space and Flight Center]
{Science and Technology division, NASA Marshall Space Flight Center, Huntsville, AL-35802, USA}

\author{Don A. Gregory}
\alsoaffiliation[Department of Physics, University of Alabama in Huntsville]
{Department of Physics and Astronomy, University of Alabama in Huntsville, Huntsville, AL-35805, USA}

\title[An \textsf{achemso} demo]
  {Silicon-Enhanced Nanocavity: From Narrow Band Color Reflector to Broadband Near-Infrared Absorber}

\abbreviations{IR,NMR,UV}

\begin{document}







\begin{abstract}
Subwavelength-scale light absorbers and reflectors have gained significant attention for their potential in photonic applications. These structures often utilize a metal-insulator-metal (MIM) architecture, similar to a Fabry-Perot nanocavity, using noble metals and dielectric or semiconductor spacers for narrow-band light absorption. In reflection mode, they function as band-stop filters, blocking specific wavelengths and reflecting others through Fabry-Perot resonance. Efficient color reflection requires asymmetric Fabry-Perot cavities, where metals with differing reflectivities and extinction coefficients enable substantial reflection for non-resonant wavelengths and near-perfect absorption at resonant ones. Unlike narrowband techniques, broadband absorption does not rely on a single resonance phenomenon. Recent developments show that integrating an asymmetric Fabry-Perot nanocavity with an anti-reflection coating achieves near-unity absorption across a broad wavelength range. This study introduces an asymmetric Fabry-Perot nanocavity with a dielectric-semiconductor-dielectric spacer, enabling near-unity color reflection. By incorporating silicon, the reflected color can be tuned with just a 5 nm thickness variation, while achieving broadband absorption over 70\% in the 800-1600 nm range. The addition of an anti-reflection coating extends broadband absorption to near unity with minimal impact on reflected color. The planar, nanopattern-free design holds promise for display technologies with better color fidelity and applications in thermal photovoltaics.

\end{abstract}

\section{Introduction}
In recent years, nanostructured color filters have garnered significant attention owing to their compact design, superior performance relative to pigment-based color dyes, and high efficiency in terms of color reflection, transmission, and absorption \cite{Kats_color_filter, shurvintonMDMD, Ghobadi_color_filter, parkvivid, o_idea_3}. These devices may become crucial in the advancement of display technology, imaging, sensing, and photovoltaics \cite{ch_5_display_1,ch_5_display_2,ch_5_display_3, dixit2024electro,ch-5_imaging_1,ch_5_imaging_2,ch_5_sensing_1,ch_5_sensing_2, yong2016narrow, wu2017ultra,ch_5_solar_1,ch_5_solar_2, wang2015titanium, wang2013perfect}. Despite their impressive capabilities in these applications, certain structural and material constraints limit the full realization of their potential. Additionally, the nano-patterning processes essential for their fabrication limit their scalability, particularly in the context of large-scale device production.

To address these challenges, lithography-free planar structures have gained attention. Such structures typically employ a metal-insulator-metal (MIM) configuration, comprising plasmonic materials or specially designed metamaterials to modulate light reflection, transmission, or absorption \cite{yoon2010transmission, o_idea_4, li2015omnidirectional, chirumamilla2016multilayer, zhong2015fully, InAs_InSb_Dixit, park2015omnidirectional}. These MIM configurations resemble Fabry-Perot nanocavities, which are known for their spectrally selective color absorption and transmission properties.

Nanoscale color filters that couple the incident wave to the fundamental Fabry-Perot resonance mode have been a hot topic recently. Fabry-Perot nanocavities (FPN) that include a dielectric or semiconductor medium sandwiched between two reflecting metal surfaces are well-known architectures for color filters \cite{mirshafieyanSi_Al, InAs_InSb_Dixit, Aydin1, Ghobadi_color_filter, P-InSb_Dixit}. One of the major reasons for the popularity of FPNs is the straightforward architecture that does not require special nanopatterning. This makes such architectures a great choice for nanoscale device applications where a large surface area is required, as in many sensing applications. Although FPNs are great contenders for color filter applications, their utility lies in preventing a resonant wavelength band from being reflected while reflecting a non-resonant wavelength band. This property makes FPNs a great filter where red-green-blue (RGB) colors are required in transmission or subtractive colors such as cyan-magenta-yellow (CMY) are required in reflection. Thus, FPNs are also known as transmissive color filters rather than reflective ones. For the creation of FPN-based color filters, noble metals such as gold, silver, and aluminum are popular choices because they provide high reflection and low absorption throughout the visible spectrum. These noble metals, when combined with a dielectric spacer that has low absorption, can be used to create perfect subwavelength transmissive color filters. In contrast to transmissive color filters, a recent innovative approach has been developed to create a color filter in reflection mode by strategically using a combination of Fabry-Perot cavities that complement each other \cite{Ghobadi_color_filter}. In this design, the second cavity absorbs the light transmitted by the first, resulting in the reflection of only the wavelength band that is not supported by either cavity. Additionally, some methods incorporate a specific metamaterial layer with noble metals and dielectrics to achieve reflective color filters. Most of these designs adopt an architecture featuring a metal-dielectric-metal configuration, where the metals have differing reflectivities and extinction coefficients \cite{shurvintonMDMD, ref_15, parkvivid, o_idea_3}. Such configurations are known as asymmetric Fabry-Perot nanocavities (AFPNs). In AFPN the top and bottom metals have different optical characteristics to make them complimentary reflectors and absorbers for different bands in the visible range. AFPN with a combination of lossy metals such as titanium (Ti), nickel (Ni), chromium (Cr), and tungsten (W) with noble metals such as gold (Au), silver (Ag), and aluminum (Al) are also useful in developing broadband absorbers \cite{ref_13,ref_15,ref_16,ref_39, zhong2016multimetal, o_idea_4}. Thus, in addition to color transmission and reflection-based devices, AFPN architecture-based nanoscale devices can be useful for broadband absorption. These devices have found potential applications in diverse fields such as biomedical optics, antenna systems, thermal emitters, and thermal photovoltaics \cite{ref_39, ref_59, ref_60, Anker_bio_sensing}. 

Although it appears to be a straightforward design, engineering an efficient broadband absorber is significantly different from designing a narrowband one. In most broadband studies, nanoscale devices incorporate patterned metallic gratings on dielectrics or plasmonic resonant structures; however, the scalability of such devices is limited by the required nanopatterning. Consequently, AFPNs with an anti-reflection (AR) coating have emerged as a promising solution to match the broadband absorption efficiency of patterned nanostructures \cite{ref_13, ref_15, parkvivid}. To achieve broadband absorption with AFPNs, metals with specific optical properties must be used. An additional AR coating on top of the AFPN structure plays a crucial role in this architecture. In some instances, AR coatings are engineered with a refractive index that lies between that of the surrounding medium and the Fabry-Perot nanocavity material. This gradation in the refractive index reduces the contrast in the refractive index at the interfaces, subsequently decreasing reflected light. AR coatings exploit interference effects, relying on destructive interference. These coatings are mostly tailored to a thickness that is one-quarter of the wavelength of the targeted light transmission. When light reflects from the coating's top surface and the interface between the coating and the cavity material, the out-of-phase nature of these reflections leads to destructive interference, effectively neutralizing the reflected waves and minimizing reflection. This determines the performance of AR coatings for enhancing the efficiency of broadband absorbers.

Focusing on the use of AFPNs as reflective color filters, recent studies present several optimal material combinations for narrowband color reflection, involving metals like platinum (Pt), bismuth (Bi), nickel (Ni), and titanium (Ti), paired with noble metals such as gold (Au), aluminum (Al), and silver (Ag) \cite{parkvivid, ref_15, Aydin1, shurvintonMDMD, o_idea_3}. These devices typically consist of two metal layers separated by a dielectric or semiconductor material such as silicon dioxide ($\text{SiO}_2$), lithium fluoride (LiF), aluminum oxide ($\text{Al}_2\text{O}_3$), or titanium dioxide ($\text{TiO}_2$). A study by Shurvinton \textit{et al.} explored an AFPN with Ag and Ti layers using $\text{SiO}_2$ as a spacer to achieve high-chroma color coatings in the visible spectrum \cite{shurvintonMDMD}. Similarly, Park \textit{et al.} conducted a study using Pt and Al with a $\text{TiO}_2$ spacer \cite{parkvivid}. Both studies leveraged the resonance properties of AFPNs to achieve color reflection.

While the structure proposed by Shurvinton \textit{et al.} achieves high chroma color coating, the ratio of the $\text{SiO}_2$ thickness to the peak reflected wavelength is notably high. For the red spectrum, this ratio stands at approximately 1/3.3, 1/1.6 for green, and 1/3.75 for blue. This elevated ratio implies that effecting a substantial shift in the resonance wavelength requires a more substantial alteration in the thickness of the dielectric spacer. The utilization of $\text{SiO}_2$, a lossless material, contributes significantly to these elevated ratios. Previous studies on narrowband absorbers have shown that replacing the spacer with a realtively lossy material can substantially decrease the spacer thickness requirement. Additionally, such structures exhibit enhanced spectral sensitivity to changes in spacer thickness \cite{Aydin1, mirshafieyanSi_Al}.

The use of a lossless material like $\text{SiO}_2$ not only results in a high spacer thickness to peak wavelength ratio but also increases the structure's sensitivity to incident angles other than normal. Under oblique illumination, the extended optical path within the cavity induces a blue shift in the resonance wavelength, a phenomenon more pronounced in reflection mode due to the lengthier optical path of light. One of the methods to achieve an angle-insensitive resonance response involves increasing the refractive index and reducing the cavity thickness. Incorporating a relatively lossy material, like a semiconductor as the cavity medium can alter the phase upon reflection at the metal-semiconductor or dielectric-semiconductor interface, allowing for Fabry-Perot resonance at the desired wavelength with thinner layers \cite{mirshafieyanSi_Al, Ghobadi_color_filter}. This adjustment would result in a minimal optical path length difference between normal and oblique incident light. A caveat of this approach is light absorption by the cavity's semiconductor layer, potentially diminishing the color filter's efficiency. This effect is particularly significant at shorter wavelengths, where the semiconductor's extinction coefficient ($\kappa$) is higher \cite{mirshafieyanSi_Al}. Therefore, substituting a lossless dielectric layer entirely with a lossy semiconductor may not be the preferred strategy.

In consideration of these factors, the present study introduces a planar, scalable, and lithography-free reflective color filter, demonstrating substantial efficiency in color reflection. The proposed architecture utilizes a Metal-Dielectric-Semiconductor-Dielectric-Metal (MDSDM) cavity structure. The layers of this cavity are sequentially composed of Ag as the bottom metal, $\text{SiO}_{2}$ as a dielectric layer, silicon (Si) as a semiconductor layer, $\text{SiO}_{2}$, and Ti as the top metal layer. The strategic placement of a Si layer, sandwiched between two $\text{SiO}_{2}$ layers, significantly reduces both the spacer thickness and the angle sensitivity of the reflector. Furthermore, the Si layer not only contributes to the effectiveness within the visible spectrum but also creates considerable broadband absorption in the near-infrared (NIR) range. This enhanced performance is attributed to the multilayer configuration of the three materials (Ag, Si, and Ti), each with distinct $\kappa$, effectively targeting regions of both the visible and NIR wavelength spectra. Silicon's higher $\kappa$ at lower wavelengths attenuates these wavelengths, while titanium and silver suppress higher wavelengths \cite{silicon_refractive, titanium_refractive, silver_refractive}. The rationale behind choosing silicon as a spacer material is twofold. Firstly, silicon is one of the most abundant elements on earth, offering cost-effectiveness compared to other semiconductor materials. Secondly, the deposition of ultrathin silicon films uses established techniques, such as radio-frequency (RF) sputtering and chemical vapor deposition (CVD). There are no inherent optical constraints preventing the substitution of silicon with other materials that exhibit a similar trend in $\kappa$ values to provide an efficient reflection phase change at the dielectric-semiconductor interface for creating color reflectors.

\section{Experimental and Simulation Methods}

\subsection{Device Fabrication}

\begin{figure}
    \centering
    \includegraphics[width = 0.7\linewidth]{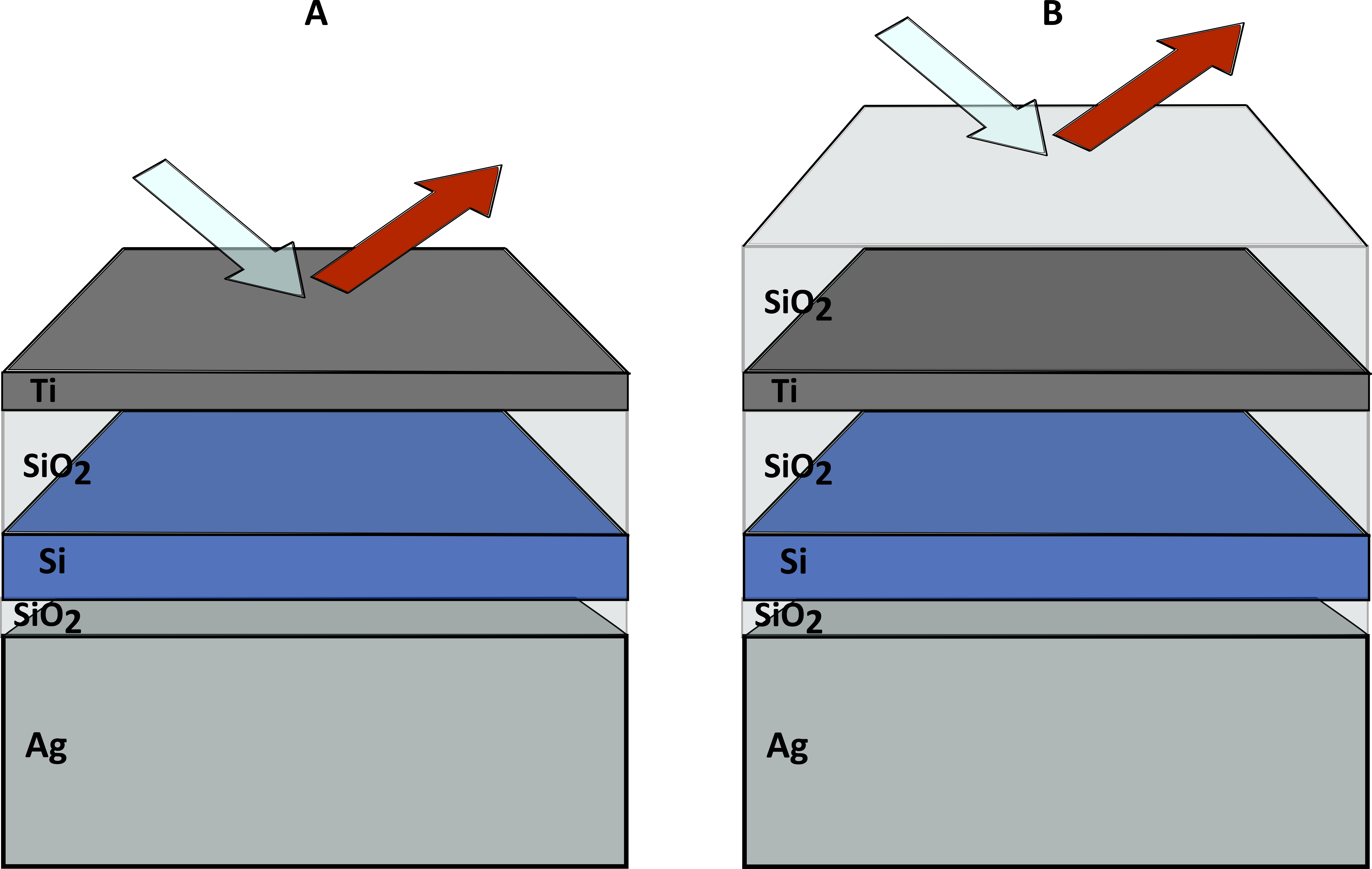}
    \caption{Schematic diagram of the proposed (A) Ti-$\text{SiO}_{2}$-Si-$\text{SiO}_{2}$-Ag and (B) $\text{SiO}_{2}$-Ti-$\text{SiO}_{2}$-Si-$\text{SiO}_{2}$-Ag asymmetric Fabry-Perot nanocavities.}
    \label{fig:AFPN schematic}
\end{figure}

Figure \ref{fig:AFPN schematic} provides schematics of the MDSDM asymmetric nanocavity structure as well as an AR-coated MDSDM architecture also referred to as MDSDMD further in the article. The MDSDM structure comprises a sequence of $\text{SiO}_{2}$-Si-$\text{SiO}_{2}$ layers, sandwiched between a Ti layer on top and a substantially thick Ag film at the bottom. The bottom Ag film, with its considerable thickness, effectively prevents light from reaching the substrate.
During the device fabrication process, a 100 nm thick Ag film was initially sputtered onto six microscope slides. This was accomplished using a Denton discovery-18 sputtering system, operating at 100 W DC power, an argon plasma pressure of 5 mTorr, a pre-deposition conditioning duration of 60 seconds, and a deposition time of 300 seconds. The deposition rate for the Ag film was determined to be 0.51 nm per second, a measurement obtained using a Wyko NT1100 white light interferometer from Veeco. The Ag deposited slides served as substrates for the subsequent deposition of a 10 nm thick $\text{SiO}_{2}$ layer, a Si layer for 5 nm, 10 nm, and 20 nm, an 80 nm $\text{SiO}_{2}$ layer, and a 10 nm thick Ti layer, all applied onto three Ag-coated substrates. The remaining three substrates (slides) received identical layering, with an additional 80 nm thick $\text{SiO}_{2}$ layer serving as an anti-reflection (AR) coating for exploration of the broadband absorption aspect of the presented AFPN. The Danton discovery-18 sputtering system has three cathodes operational in DC/RF mode. Ag, $\text{SiO}_{2}$, and Si targets were applied to these cathodes, with metal cover plates protecting the targets from contamination. The deposition of Ag and first $\text{SiO}_{2}$ layers was performed without venting the chamber to avoid the oxidation of the Ag layer. Similarly, the deposition of Si and second $\text{SiO}_{2}$ layers on each sample individually was performed without venting the chamber to avoid oxidation of Si. These depositions resulted in six multilayer thin film structures, with three of them representing a Metal-Dielectric-Semiconductor-Dielectric-Metal structure, while the rest with Metal-Dielectric-Semiconductor-Dielectric-Metal-Dielectric (MDSDMD) structures. The MDSDM structures are Ag (100 nm) - $\text{SiO}_{2}$ (10 nm) - Si (5 nm, 10 nm, and 20 nm) - $\text{SiO}_{2}$ (80 nm) - Ti (10 nm), and the MDSDMD structures are Ag (100 nm) - $\text{SiO}_{2}$ (10 nm) - Si (5 nm, 10 nm, and 20 nm) - $\text{SiO}_{2}$ (80 nm) - Ti (10 nm) - $\text{SiO}_{2}$ (80 nm). The selection of these particular thickness values for each layer is discussed in the structure modeling section of this article. The power settings, argon plasma pressure, and pre-sputtering times for all layers except for Si were consistent with those used for Ag deposition. For Si the sputtering power was set to 200W RF with 4.85 m Torr argon pressure. The calibrated deposition rates for $\text{SiO}_{2}$, Si, and Ti, as measured by the white light interferometer, were 0.02 nm per second, 0.34 nm per second, and 0.15 nm per second, respectively.

\subsection{Simulations and Measurements}

The modeling of the MDSDM and MDSDMD structures was performed using the transfer matrix method (TMM) \cite{Pochi_TMM}. In addition to this modeling, optical simulations were conducted using commercial finite-difference-time-domain (FDTD) software, Ansys Lumerical FDTD \cite{Lumerical_fdtd}. These simulations were three-dimensional (3D) in nature. A plane-wave source was employed, spanning the desired wavelength range of 400-1600 nm. For the boundary conditions, periodic settings were applied in the lateral (x and y) directions, while the vertical (z) direction (AFPN thickness) was configured with a perfectly matched layer (PML) to effectively absorb outgoing waves and prevent reflections.

The optical reflectivity within the visible spectrum of the sample was experimentally measured using a broadband halogen light source in conjunction with an optical spectrometer (Ocean Optics Flame Miniature Spectrometer). For the assessment of normal incidence light reflection from the structure, a normal reflection/backscattering probe, also provided by Ocean Optics, was employed. In all of the measurements, the obtained reflection values are normalized with the reflection data from a thick Ag-coated sample. The real and imaginary parts of the refractive index of the Ag, $\text{SiO}_{2}$, Si, and Ti within the visible wavelength range were determined using a J.A. Woollam Alpha-SE Spectroscopic Ellipsometer. For the near-infrared wavelength range, the refractive index data for Ag, $\text{SiO}_{2}$, Si, and Ti thin films were obtained from the studies by Ciesielski \textit{et al.} \cite{silver_refractive}, Gao \textit{et al.} \cite{gao2012exploitation}, Pierce \textit{et al.} \cite{silicon_refractive}, and Palm \textit{et al.} \cite{titanium_refractive}.

\subsection{Structure Modeling}

To achieve effective color reflectors and broadband absorbers, the selection of materials for each layer within the stack plays a crucial role. The application of the transfer matrix method (TMM) modeling approach assists in determining the optimal thickness for each layer, thereby achieving the desired reflectance in the visible spectrum and broadband absorption in the near-infrared range.

In the present study, Ag and $\text{SiO}_{2}$ coatings are the chosen materials. Ag is preferred as the thick bottom reflector metal due to its high reflectivity in the visible range, making it a suitable choice for developing a color reflector in reflection mode. $\text{SiO}_{2}$, characterized as a lossless material, enhances multiple beam reflections within the Fabry-Perot cavity which helps increase light absorption in the metal layers.

Above the $\text{SiO}_{2}$ coating, Si is introduced as the third layer as a material with a relatively higher extinction coefficient, especially when compared to $\text{SiO}_{2}$. As previously discussed, the higher extinction coefficient of Si in the lower range of visible wavelengths aids in suppressing reflectance within that spectral region. Additionally, the inclusion of Si reduces the angular sensitivity of incident waves by decreasing the overall spacer thickness (especially in the reflection mode). To mitigate a relatively higher reflection phase change at the metal-semiconductor (Ti-Si) interface, an $\text{SiO}_{2}$ layer is added atop the Si film. The uppermost layer consists of Ti to establish an asymmetric Fabry-Perot cavity. A schematic structure for the AFPN is presented in fig. \ref{fig:AFPN schematic}.

\begin{figure}
    \centering
    \includegraphics[width = \linewidth]{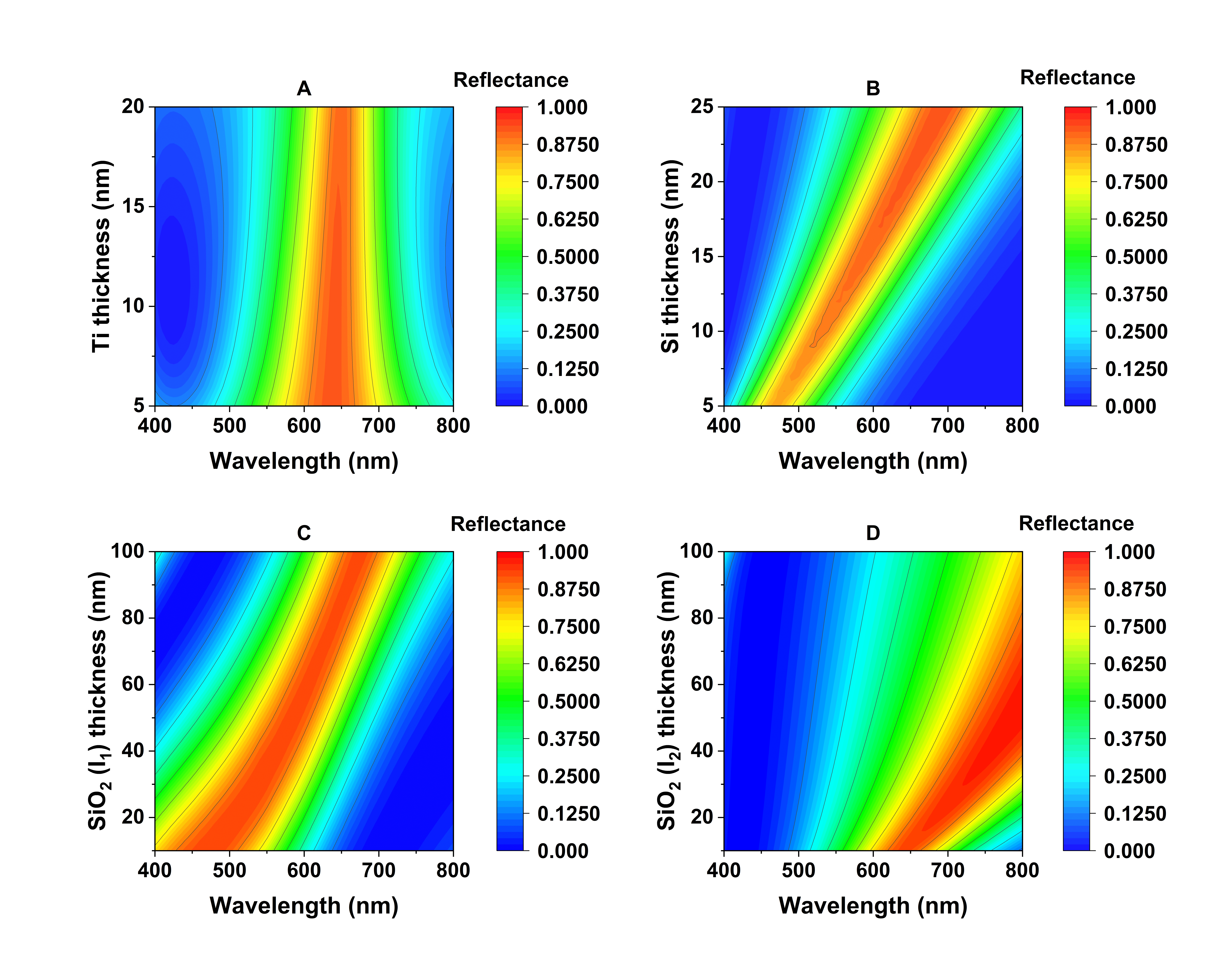}
    \caption[The contour plots of the reflectance spectra.]{The contour plots of the reflectance spectra for different (A) Ti thickness (B) Si thickness (C) top $\text{SiO}_{2}$ thickness, and (D) bottom $\text{SiO}_{2}$ thickness.}
    \label{thickness_contour}
\end{figure}

 The initial phase of the study involved calculating the round trip phase delay for the Ti-$\text{SiO}_{2}$-Si-$\text{SiO}_{2}$-Ag structure, targeting a reflection wavelength of 600 nm. The purpose was to fine-tune the thickness of the $\text{SiO}_{2}$-Si-$\text{SiO}_{2}$ combination to induce destructive interference at wavelengths outside the vicinity of 600 nm. Numerical simulations employing the TMM were utilised, incorporating the refractive indices for each layer. The thicknesses of the Ag, bottom $\text{SiO}_{2}$, top $\text{SiO}_{2}$, and Ti layers were fixed at 100 nm, 40 nm, 40 nm, and 10 nm respectively, while the Si layer's thickness was varied to determine the optimal value. The reflection contour plot for varying Si thicknesses within the Ti-$\text{SiO}_{2}$-Si-$\text{SiO}_{2}$-Ag MDSDM cavity configuration is presented in fig. \ref{thickness_contour}(B). It is evident that a minor alteration in Si thickness results in a significant shift in the peak reflection wavelength.

Variations in the Ti layer thickness (fig. \ref{thickness_contour}(A)) demonstrates that while such adjustments do not significantly shift the peak reflected wavelength, they contribute to narrowing the bandwidth of the reflected spectrum, thereby enhancing color purity. The choice of Ti thickness must account for its absorptive properties, ensuring the thickness does not exceed the metal's optical penetration depth. Adjustments to the top $\text{SiO}_{2}$ layer thickness (fig. \ref{thickness_contour}(C)) result in a notable shift in the peak reflected wavelength across the visible spectrum. However, alterations to the thickness of the bottom $\text{SiO}_{2}$ layer (fig. \ref{thickness_contour}(D)) are less effective.

\begin{figure}
    \centering
    \includegraphics[width = 0.65\linewidth]{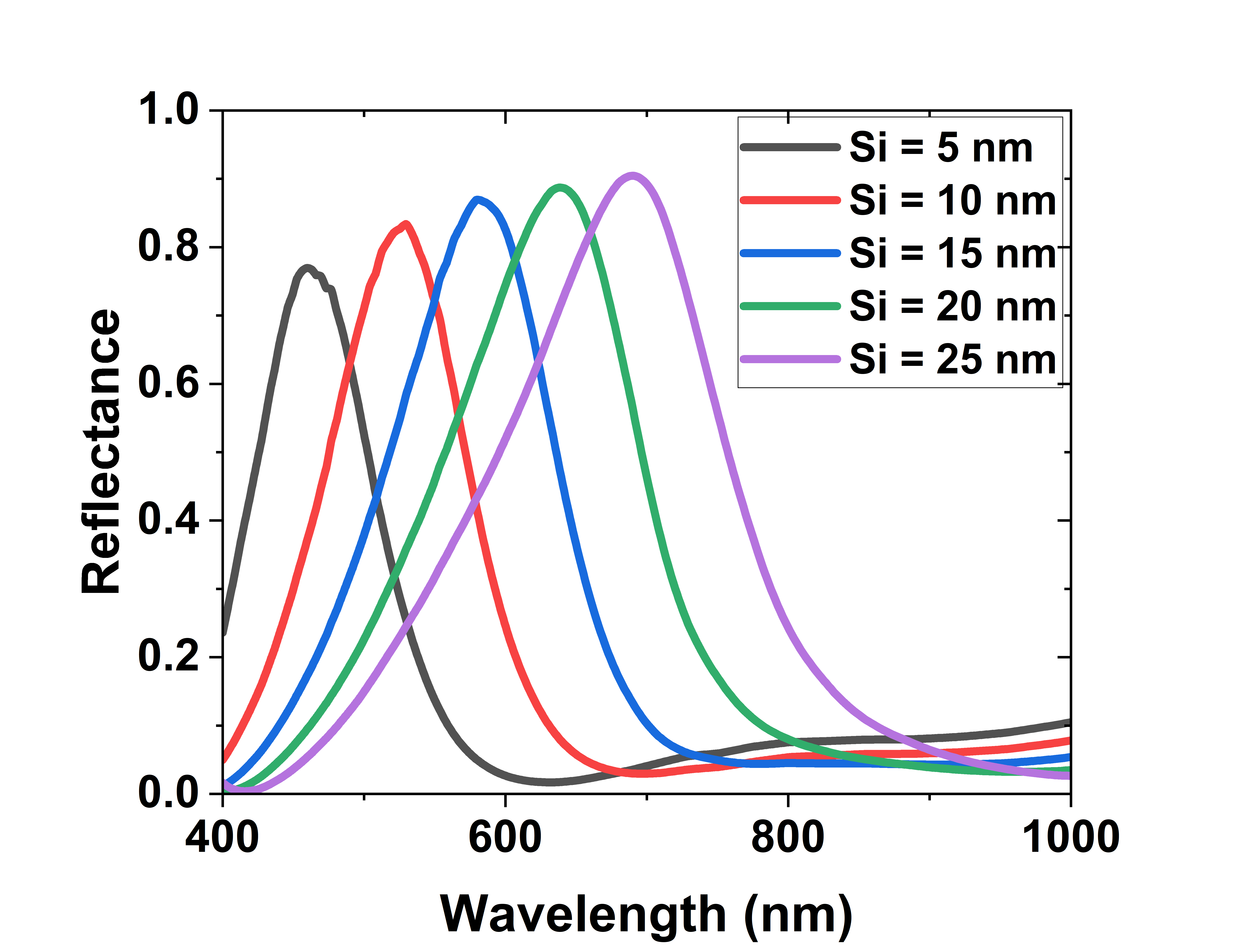}
    \caption{Reflectance spectra for Ti (10 nm)-$\text{SiO}_{2}$ (80 nm)-Si (5-25 nm)-$\text{SiO}_{2}$ (10 nm) - Ag (100 nm) at normal incidence.  }
    \label{fig:AFPN refelctance}
\end{figure}

\begin{figure}
    \centering
    \includegraphics[width = 0.85\linewidth]{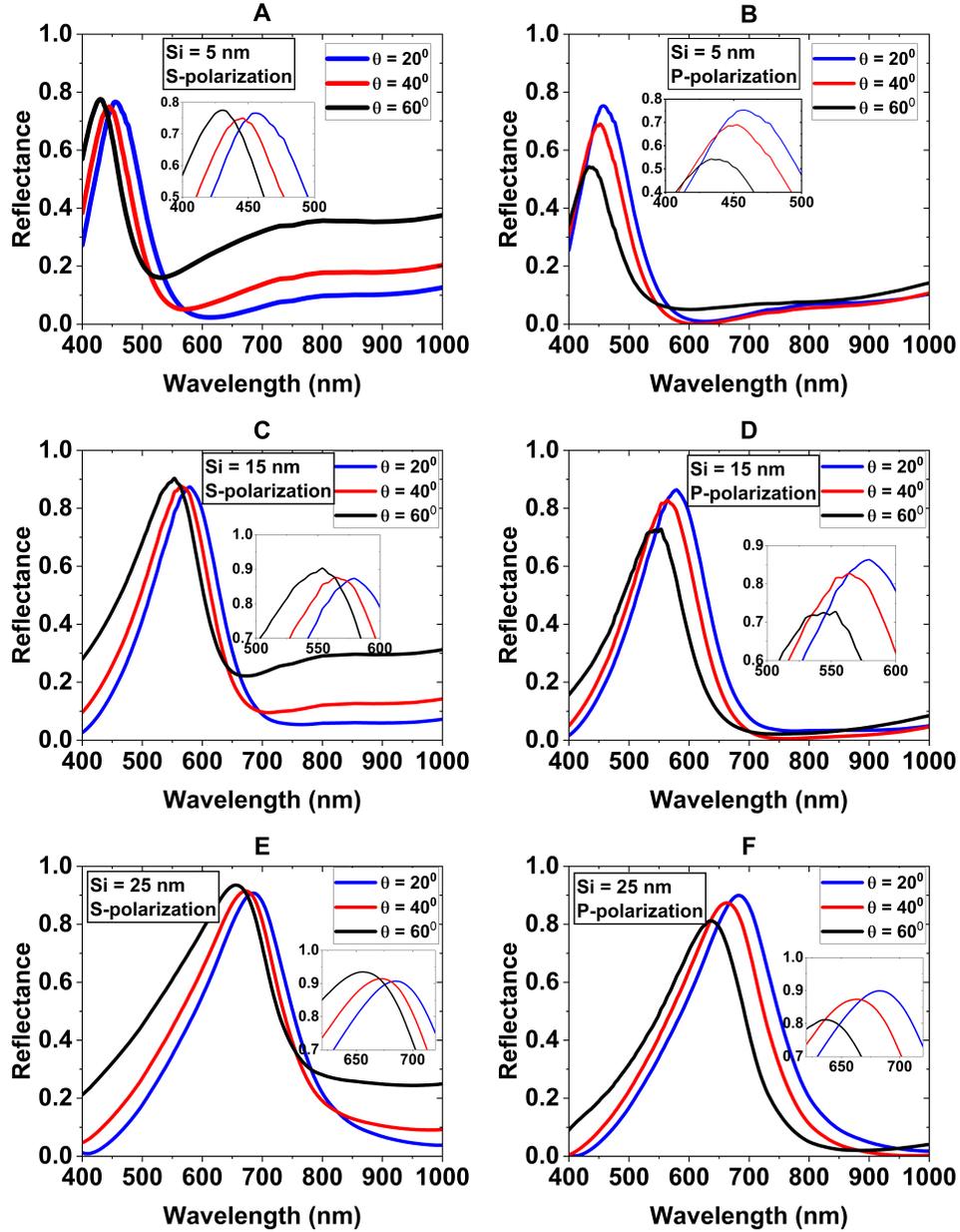}
    \caption{Reflectance spectra for (A) Ti (10 nm) - $\text{SiO}_{2}$ (80 nm) - Si (5 nm) - $\text{SiO}_{2}$ (10 nm) - Ag (100 nm) (C) Ti (10 nm) - $\text{SiO}_{2}$ (80 nm) -Si (15 nm) - $\text{SiO}_{2}$ (10 nm) - Ag (100 nm), and (E) Ti (10 nm) - $\text{SiO}_{2}$ (80 nm) - Si (5 nm) - $\text{SiO}_{2}$ (10 nm) - Ag (100 nm) for S-polarized incident wave at different incident angles. Reflectance spectra for (B) Ti (10 nm) - $\text{SiO}_{2}$ (80 nm) - Si (5 nm) - $\text{SiO}_{2}$ (10 nm) - Ag (100 nm), (D) Ti (10 nm) - $\text{SiO}_{2}$ (80 nm) - Si (15 nm) - $\text{SiO}_{2}$ (10 nm) - Ag (100 nm), and (F) Ti (10 nm) - $\text{SiO}_{2}$ (80 nm) - Si (25 nm) - $\text{SiO}_{2}$ (10 nm) - Ag (100 nm) for P-polarized incident wave at different incident angles.}
    \label{fig:5nm Si SP_1}
\end{figure}

To develop a silicon-enhanced AFPN color reflector, the top Ti layer's thickness was set at 10 nm. This parameter allows Ti to function as a layer with partial reflectivity and transmissivity and represents an optimal thickness for thin film deposition, avoiding the complexities associated with depositing ultra-thin nanofilms (3-7 nm). The top $\text{SiO}_{2}$ layer was assigned a thickness of 80 nm, achieving a narrow peak reflection band at greater thicknesses (fig. \ref{thickness_contour}(C)). The bottom $\text{SiO}_{2}$ layer's thickness was chosen to be 10 nm, selected for producing a narrow peak reflection band (fig. \ref{thickness_contour}(D)). To achieve a spectral transition from blue to red, as illustrated in Figure \ref{fig:AFPN refelctance}, the thickness of the silicon layer was methodically varied between 5 nm and 25 nm, in increments of 5 nm. Figure \ref{fig:5nm Si SP_1} illustrates the modeled reflectance spectra for AFPNs under S and P-polarized incident waves. These spectra are analyzed at incident angles of 
$20^0, 40^0, \text{and}\ 60^0$, corresponding to silicon thicknesses of 5, 15, and 25 nm, respectively. 

\begin{figure}
    \centering
    \includegraphics[width = \linewidth]{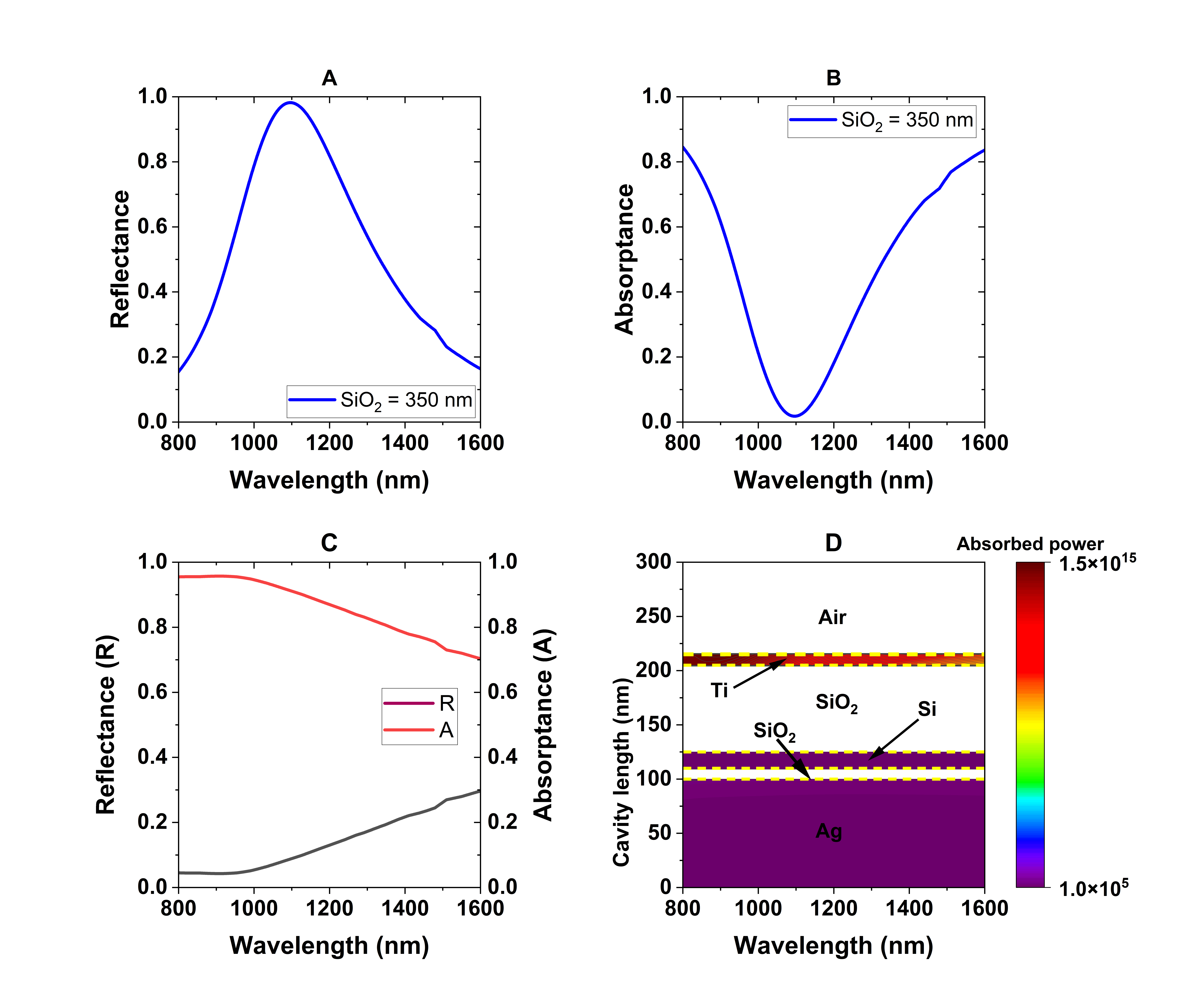}
    \caption[Reflectance and absorptance spectrum for Ti (10 nm) - $\text{SiO}_{2}$ (350 nm) - Ag (100 nm) AFPN.]{(A) Reflectance and (B) Absorptance spectrum for Ti (10 nm) - $\text{SiO}_{2}$ (350 nm) - Ag (100 nm) AFPN for part of the NIR wavelength range. (C) Reflectance and Absorptance spectrum, and (D) Absorbed power contour plot for Ti (10 nm) - $\text{SiO}_{2}$ (80 nm) - Si (15 nm) - $\text{SiO}_{2}$ (10 nm) - Ag (100 nm) AFPN for part of the NIR wavelength range.}
    \label{fig:BB Shurviton}
\end{figure}

Focusing on the near-infrared (NIR) spectrum, the implementation of silicon as a spacer medium in the cavity is observed to limit NIR wavelength reflection to below 30\% for wavelengths up to 1600 nm. Figures \ref{fig:BB Shurviton}(A) and (B) depict the reflectance and absorptance in the NIR region for the Ti-$\text{SiO}_{2}$-Ag, Metal-Dielectric-Metal (MDM) cavity with 350 nm thick $\text{SiO}_{2}$ layer (represented as green in the RGB colour reflector) developed by Shurvinton \textit{et al.} with a sole purpose of high-chroma color reflection.\cite{shurvintonMDMD} In Figure \ref{fig:BB Shurviton}(B), the absorptance (A) is governed by the relationship A+R = 1-T, where R is reflectance and T is transmittance. Given that the cavity operates in reflection mode with zero transmittance, the relationship simplifies to A = 1-R. Figure \ref{fig:BB Shurviton}(C) presents the R and A plots for the Ti(10 nm)-$\text{SiO}_{2}$(80 nm)-Si(15 nm)-$\text{SiO}_{2}$(10 nm)-Ag(100 nm) cavity in the NIR range up to 1600 nm. The relationship between A and R is consistent with the previously mentioned cavity.

A cavity comprised solely of $\text{SiO}_{2}$  as the spacer layer extends resonance modes into the NIR range. In contrast, the $\text{SiO}_{2}$-Si-$\text{SiO}_{2}$  cavity suppresses NIR range reflection more effectively. The layered arrangement of Ti, Si, and Ag is critical here, where the incorporation of a central Si layer bifurcates the cavity into Metal-Dielectric-Semiconductor (MDS) and Semiconductor-Dielectric-Metal (SDM) configurations. These two entities decrease the reflectance significantly within the 800-1600 nm wavelength range. The power absorption by each layer of the Ti-$\text{SiO}_{2}$-Si-$\text{SiO}_{2}$-Ag cavity, computed using the Finite-Difference Time-Domain (FDTD) method, is shown in Figure \ref{fig:BB Shurviton}(D). The Ti layer absorbs the most power, with the Si and Ag layers contributing to this effect, while $\text{SiO}_{2}$, as a lossless material, does not absorb power in the range observed for the metal and semiconductor layers. Consequently, the inclusion of Si in the cavity significantly reduces reflected light by enhancing absorption in the Ti-$\text{SiO}_{2}$-Si section, and transmits a reduced amount to the reflective Ag layer of the Si-$\text{SiO}_{2}$-Ag section, thereby creating a broadband absorber in the NIR range.
       
To further enhance absorption and extend the coverage across a broader range of the near-infrared (NIR) spectrum, an anti-reflective (AR) coating can be applied to the top of the Ti (10 nm) - $\text{SiO}_{2}$(80 nm) - Si (15 nm) - $\text{SiO}_{2}$(10 nm) - Ag (100 nm) AFPN structure. Suitable dielectric materials for the AR coating include $\text{SiO}_{2}$, aluminum oxide ($\text{Al}_{2}\text{0}_{3}$), and zinc oxide (ZnO).\cite{parkvivid, shurvintonMDMD, Ghobadi_color_filter} In this research, $\text{SiO}_{2}$ was chosen for the AR coating to limit the number of different materials used in the single AR coated-AFPN structure. An optimized thickness of the AR coating can lower reflectance by restricting the reflection of wavelengths that produce destructive interference due to round-trip phase delay at the air-dielectric boundary.

\begin{figure}
    \centering
    \includegraphics[width = \linewidth]{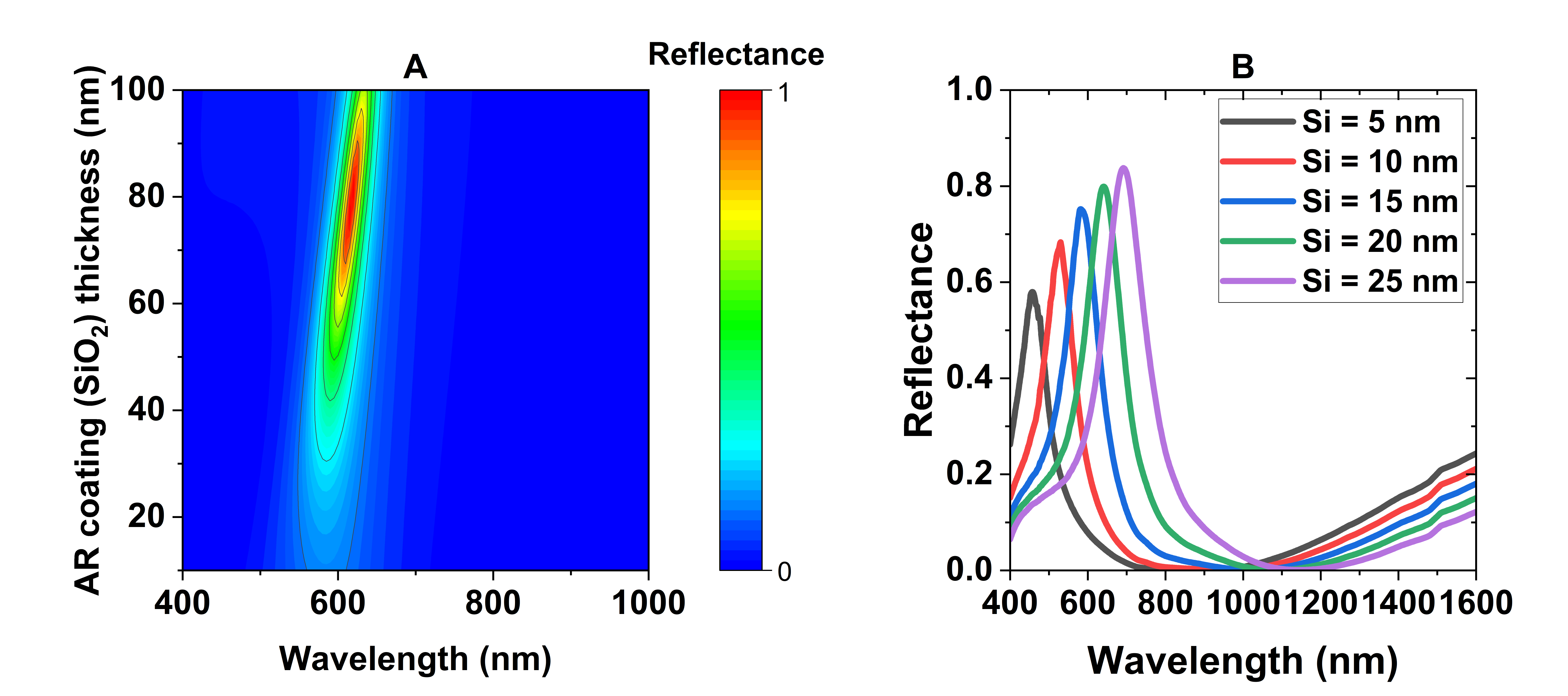}
    \caption[Reflectance contour plot as a function of AR coating.]{(A) Reflectance contour plot as a function of AR coating ($\text{SiO}_{2}$) thickness for AR coating - Ti (10 nm) - $\text{SiO}_{2}$ (80 nm) - Si (15 nm) - $\text{SiO}_{2}$ (10 nm) - Ag (100 nm) structure. (B) Reflectance spectra for AR coating ($\text{SiO}_{2}$) (80 nm) - Ti (10 nm) - $\text{SiO}_{2}$ (80 nm) - Si (15 nm) - $\text{SiO}_{2}$ (10 nm) - Ag (100 nm) AFPN for visible and NIR wavelength range.}
    \label{fig:AR coating contour}
\end{figure}

The proposed AFPN structure is not only aimed to serve as a broadband absorber in the NIR but also as a color reflector in the visible range. Therefore, determining the $\text{SiO}_{2}$ AR coating thickness should also take into account its influence on resonance modes within the visible spectrum. Figure \ref{fig:AR coating contour}(A) displays a reflectance contour plot for the Ti-$\text{SiO}_{2}$-Si-$\text{SiO}_{2}$-Ag AFPN with a 15 nm Si layer and varying top $\text{SiO}_{2}$ (AR coating) thicknesses. A thickness range of 70-90 nm for the AR coating appears to be optimal for both color reflection and broadband absorption in the AFPN. As shown in Figure \ref{fig:AR coating contour}(B), the reflectance in certain NIR wavelength regions can be reduced significantly, approaching zero, so that the AR-coated AFPN is an effective broadband absorber in the NIR spectrum.

\section{Results and discussion}

\begin{figure}
    \centering
    \includegraphics[width = 1\textwidth]{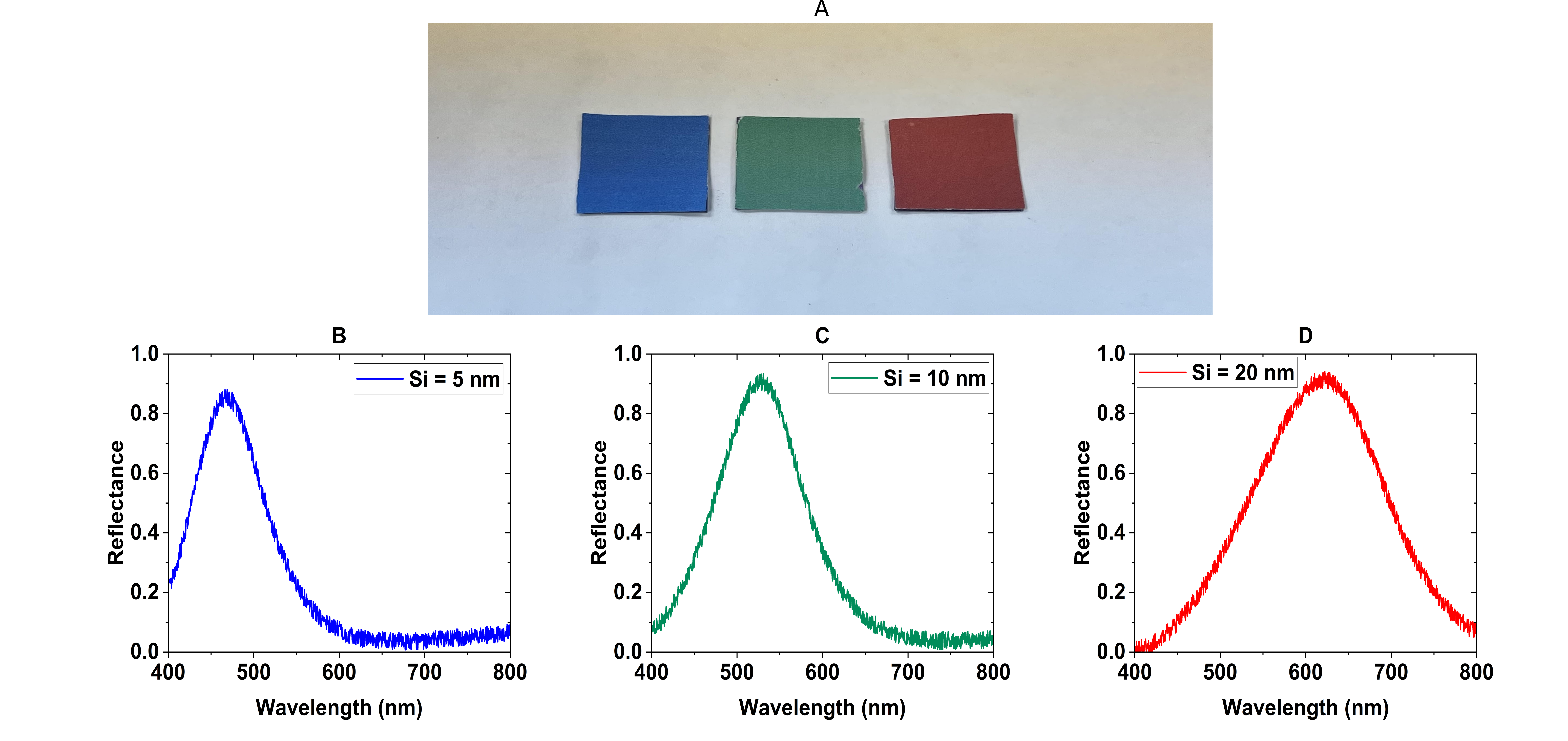}
    \caption[The optical image of the fabricated MDSDM designs operating in different wavelengths.]{(A) The optical image of the fabricated MDSDM samples with semiconductor (Si) thickness of 5 nm (blue film), 10 nm (green film), and 20 nm (orange film). Reflectance spectra of the sample with (B) 5 nm thick Si, (C) 10 nm thick Si, and (D) 20 nm thick Si.}
    \label{fig: captured image}
\end{figure}

The device under discussion, shown in fig.\ref{fig:AFPN schematic}, is based on an asymmetric Fabry-Pérot (F-P) nanocavity (AFPN), incorporating a sequential arrangement of $\text{SiO}_{2}$-Si-$\text{SiO}_{2}$ layers, acting as a spacer, situated between Ag and Ti layers. Figure \ref{fig: captured image}(A) is a picture of the prepared thin film samples for the intended color reflection and NIR absorption. A 100 nm thick Ag layer forms the base of the structure to ensure high reflection across the visible spectrum and prevent transmission through the device. The resonance wavelength of the F-P nanocavity is determined by the cavity's thickness (d), and the spectral response is fine-tuned by varying the Si layer's thickness within the cavity: 5 nm, 10 nm, and 20 nm. The optical characteristics of the Metal-Dielectric-Semiconductor-Dielectric-Metal (MDSDM) stack were investigated by measuring reflection profiles at normal incidence, using a broadband halogen light source in conjunction with an Ocean Optics FLAME miniature spectrometer and a reflection/backscattering probe. Figures \ref{fig: captured image}(B), (C), and (D) represent the experimentally measured reflectance spectra of the fabricated devices. Here, slight variations are observed in the measured reflectance spectra compared to the calculated spectra (fig. \ref{fig:AFPN refelctance}). The broadening of the non-resonant peaks indicates variations in the number density and damping factor within the silicon layer of the deposited samples compared to ideal calculations \cite{InAs_InSb_Dixit, dixit2024electro}. As described in the simulation and measurement section, a planar, thick Ag layer serves as a reference for normalizing the measured reflectance spectra. These results demonstrate that the design concept, which relies on integrating a relatively lossy material within the cavity layer, can offer a straightforward, lithography-free route for creating efficient reflective RGB color filters. This architecture differs from other Fabry-Perot-based simple MIM designs. While other designs absorb only a narrow frequency range and reflect the rest of the spectrum, this one serves as a narrowband reflective filter.

\begin{figure}
    \centering
    \includegraphics[width = 0.7\linewidth]{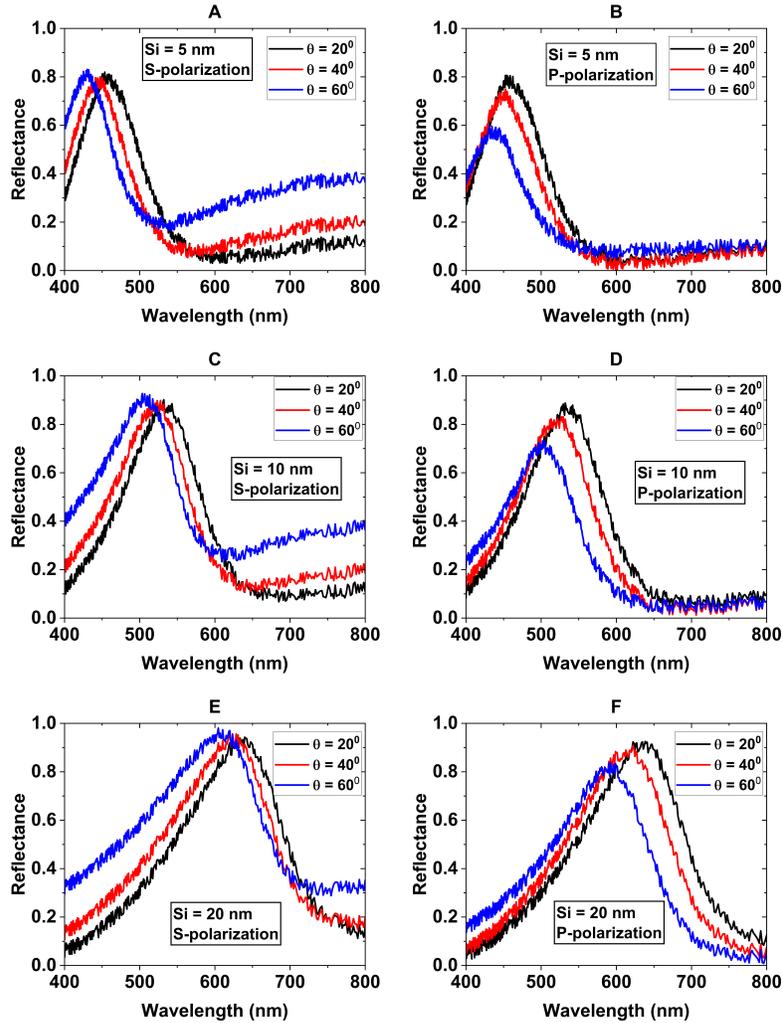}
    \caption[Blue colored filters under
different incidence angles for S-polarization.]{Reflectance spectra of the sample with S-polarized light for (A) 5 nm thick Si, (C) 10 nm thick Si, and (E) 20 nm thick Si at different angles. Reflectance spectra of the sample with P-polarized light for (B) 5 nm thick Si, (D) 10 nm thick Si, and (F) 20 nm thick Si at different angles. }
    \label{fig:experimental_1_s}
\end{figure}

An important factor limiting the practical utility of a color filter is its angular response under various angles of incident light. Figure \ref{fig:experimental_1_s} illustrates the angular response of the fabricated thin film reflectors for S and P light polarizations, with incident angles $\theta$ = 20°, 40°, and 60°. For all samples, the peak position undergoes a blue shift for both polarizations. As the angle of incident light widens, leading to light refraction at different interfaces, the optical path lengthens, resulting in the expected blue shift for the samples when illuminated with angled light. In this planar configuration, varying Fresnel reflection coefficients for S and P polarizations result in predictable angular responses.

\begin{figure}
    \centering
    \includegraphics[width = \linewidth]{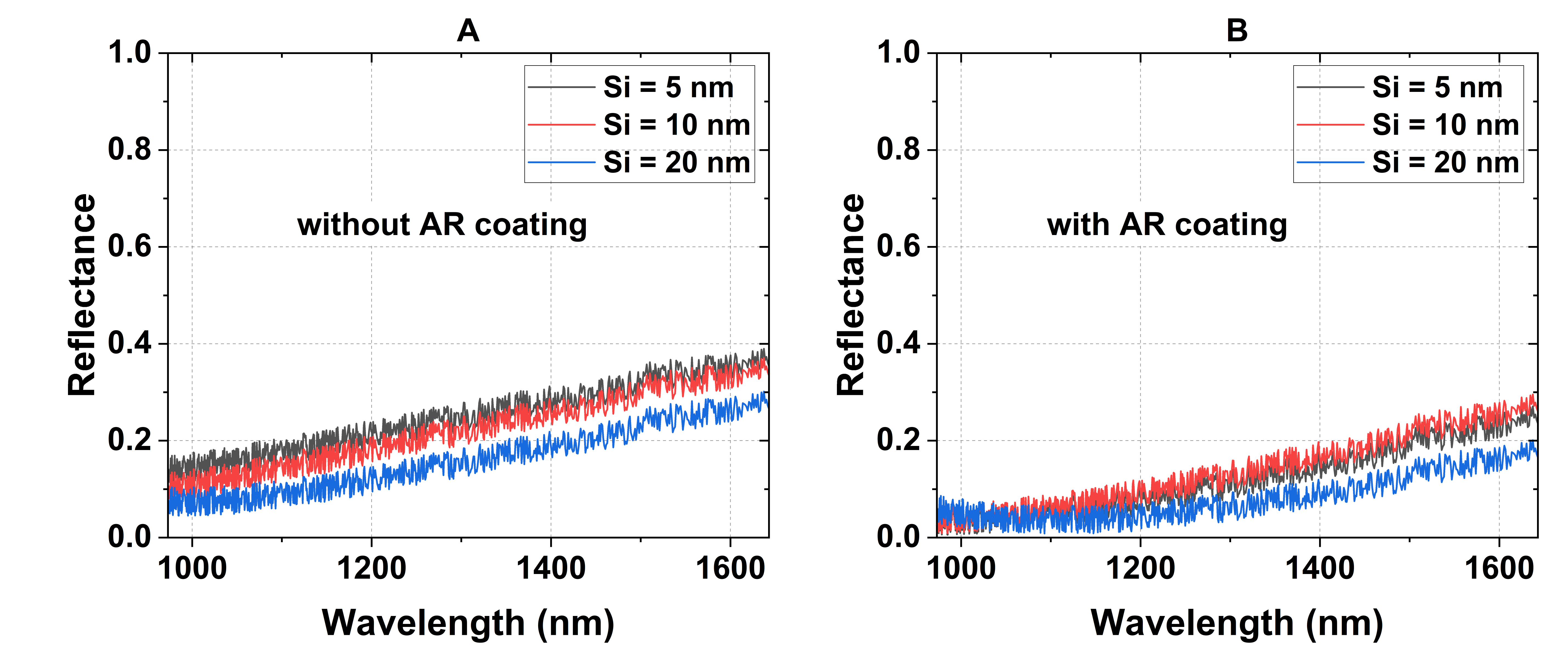}
    \caption[Measured NIR reflectance spectra.]{Measured NIR reflectance spectra for (A) Ti (10 nm)- $\text{SiO}_{2}$ (80 nm) - Si (5, 10, and 20 nm) - $\text{SiO}_{2}$ (10 nm) - Ag (100 nm) (B) $\text{SiO}_{2}$ (80 nm) - Ti (10 nm)- $\text{SiO}_{2}$ (80 nm) - Si (5, 10, and 20 nm) - $\text{SiO}_{2}$ (10 nm) - Ag (100 nm).}
    \label{fig:experimental BB}
\end{figure}

To evaluate the broadband response of the constructed sample in the near-infrared (NIR) region, a broadband visible to NIR light source was employed along with the Ocean Optics Flame NIR miniature spectrometer. Measurements were conducted exclusively at normal incidence and are depicted in fig. \ref{fig:experimental BB} The spectral range of the NIR spectrometer extends from 970 to 1640 nm, whereas the visible spectrometer spans 200 to 830 nm. This results in a data collection gap in the 830-970 nm wavelength range. However, this absence of data does not impact the research objectives, as the experimental investigation of fabricated AFPNs as color reflectors is limited to the 400-800 nm range, and the broadband NIR absorption is expected to be substantial within 900-1250 nm (as shown in figs. \ref{fig:BB Shurviton}(C)). The experimental results agree well with the predicted response of the AFPN and the AR-coated AFPN in the NIR spectrum, as illustrated in fig. \ref{fig:experimental BB}.

\section{Conclusions}
This study investigates the optical response of an asymmetric Fabry-Perot nanocavity (AFPN), that incorporates a series of dielectric-semiconductor-dielectric layers as a spacer medium in the design of deep sub-wavelength resonant cavities. Employing a transfer matrix method (TMM)-based modeling approach, the optical response of the nanocavity in both the visible and near-infrared (NIR) ranges was initially determined, focusing on the use of a semiconductor to augment the dielectric material in the cavity spacer. Unlike other designs that utilize dielectric materials as cavity spacers and have problems with low spectral shift sensitivity and high angular sensitivity issues, the inclusion of silicon within the cavity spacer noticeably enhances the spectral shift sensitivity to changes in Si thickness and significantly reduces the angular sensitivity of the color reflector compared to AFPNs using standard dielectric spacers. Furthermore, the structure was also predicted to function as a broadband absorber in the NIR range.

Numerical simulations were then conducted to identify optimal geometries for achieving peak performance. Characterization of the fabricated samples revealed that the Metal-Dielectric-Semiconductor-Dielectric-Metal (MDSDM) cavities exhibit both broadband absorption of the near-infrared (NIR) and narrowband color reflection in the visible range. It was also observed that the spectral position of the reflection peak could be tuned by varying the thickness of the silicon layer. The planar structure demonstrated over 80\% absorption of incident light across a broadband range from 800 nm to 1300 nm while simultaneously reflecting a narrow band of visible light. The NIR absorption was extended up to 1600 nm by adding an anti-reflection coating less than 100 nm thick on top of the AFPN, without significantly affecting the color reflection performance.

In conclusion, the findings of this study offer a cost-effective method for creating planar nanoscale optical structures capable of serving as RGB color filters in reflection mode, while simultaneously acting as broadband absorbers in the NIR range. This dual functionality makes the design efficient for color filtering and sensing applications in the visible spectrum, as well as for photovoltaic and communication systems operating in the NIR wavelength range.

\section{Data availability}

The datasets calculated, collected, and/or analyzed during the current study are available from the corresponding author upon reasonable request.

\section{Author Contributions}
K.P.D. and D.A.G. equally contributed to this work.

\bibliography{acs-achemso}

\end{document}